\documentstyle[12pt]{article}

\begin{document}

\title{The use of the algebraic programming in teaching general relativity
\footnote{accepted for publication in Computing in Science and Engineering, 
2000}}
\author{Florin A. Ghergu and Dumitru N. Vulcanov \\
{\small \textit{Theoretical and Computational Physics Department}}\\
{\small \textit{The West University of Timi\c soara}}\\
{\small \textit{Bv. V. P\^ arvan No. 4, 1900 Timi\c soara, Rom\^ ania}}\\
{\small \textit{E-mail : \verb|florin/vulcan@mitica.physics.uvt.ro|}}}
\date{}

\maketitle

\vspace*{5mm}

\textbf{Abstract}

{\small \noindent The article presents some aspects on the use of computer
in teaching general relativity for undergraduate students with some
experience in computer manipulation. The article presents
some simple algebraic programming (in REDUCE+EXCALC package) procedures for
obtaining  and the study of some exact  solutions of the Einstein equations in order to
convince a dedicated student in general relativity about the utility of a
computer algebra system.} 

\section{Introduction}

Teaching general relativity is a very difficult task not only for the
"teacher" but also for the students. But why ?

General relativity (GR) is not only a theory of gravity; it is a theory of
the structure of space and time, and hence a theory of the dynamics of the
universe in its entirety. Thus, the theory is a vast edifice of \textbf{pure
geometry}, indisputably elegant, but of a great \textbf{mathematical
difficulty} especially for undergraduate students (among others...). But why
undergraduate students? Because, in our department, (considering GR a
necessity for the medium graduate physics people) we make efforts to
introduce a course of GR at the III-rd year level. After some years of
experience in teaching GR at graduate students, specialized in theoretical
physics, we can say that after some weeks of introducing the must important
tools of differential geometry, starting with the physical problems of GR,
an important part of our students were almost discouraged. At an
undergraduate level, with unspecialized students, this risk is much greater.

Thus, algebraic programming systems (like REDUCE - \cite{1}) 
which contain differential
geometry packages can become a very important tool for surpassing these
difficulties. With the computer, the student can learn very fast, and in an
attractive manner, the important notions of differential geometry, tensor
calculus and, of course, the exterior calculus (with EXCALC for REDUCE, for
example - \cite{2}). As an example, he can make, after some simple computer
manipulation, the long and unattractive (for the "manual" student)
calculations from the Riemannian geometry, with metrics and Christtoffel
symbols, etc. Thus, the first  section of this article
illustrates how we can use the EXCALC package in teaching Riemannian
geometry.

But having such a powerful tool for calculus in differential geometry, we
can try to introduce some aspects of GR using computer algebra systems. For
example, the Schwarzschild solution can be easy obtained on the
computer, and also we can generate the Reissner-Nordstrom solution as exact solutions
of the Einstein equations.
These aspects are presented in the second  section of the article.

The last section of the article  presents the use of computer algebra in
finding and treating other exact homogeneous solutions of the Einstein equations 
with cosmological constant (de Sitter and anti de Sitter metrics more precisely).

As a conclusion, we consider the use of computer  as an
important tool for teaching general relativity. During the last two years we
have experienced several packages of procedures, (in REDUCE + EXCALC for
algebraic programming and in Mathematica or Maple for graphic visualizations) 
which
fulfill this purpose. Even when the students were real beginners in computer
manipulation we have obtained visible good results, in approaching several
topics of differential geometry and of course, in general relativity.

\section{Differential geometry in EXCALC}

The program EXCALC (\cite{2}) 
is completely embedded in REDUCE, thus all features
and facilities of REDUCE are available in a calculation.

EXCALC is designed for easy use by all who are familiar (or want to became)
with the calculus of Modern Differential Geometry. The program is currently
able to handle scalar-valued exterior forms, vectors and operations between
them, as well as non-scalar valued forms (indexed forms).

Geometrical objects like \textbf{exterior forms} or \textbf{vectors} are
introduced to the system by declaration commands; therefore zero-forms
(functions) must also be declared. Also, specific operations with geometric
objects are available in EXCALC like : \textbf{exterior multiplication}
between exterior forms (carried out with the nary infix operator $\wedge$
(wedge)), \textbf{partial differentiation} (is denoted by the operator @), 
\textbf{exterior differentiation} of exterior forms (carried out by the
operator d), the \textbf{inner product} between a vector and an exterior
form (represented by the diphthong $\_\mid$ (underscore or-bar)), the 
\textbf{Lie derivative} can be taken between a vector and an exterior form
or between two vectors (represented by the infix operator $\mid \_$), the 
\textbf{Hodge-* duality} operator (maps an exterior form of degree K to an
exterior form of degree N-K, where N is the dimension of the space). It is
possible to declare an indexed quantity completely antisymmetric or
completely symmetric. Some examples :
\begin{verbatim}
      PFORM U=1,V=1,W=K;   %declaration of some forms
      (3*U-A*W)^(W+5*V)^U;
      A*(5*U^V^W - U^W^W)
      @(SIN X,X);         % partial differentiation
      COS(X)
 
      PFORM X=0,Y=K,Z=M;
      D(X * Y);           % exterior differentiation of a
      X*d Y + d X^Y       % product of two forms
      D(X*Y^Z);
            K
      ( - 1) *X*Y^d Z  + X*d Y^Z + d X^Y^Z
 
      PFORM X=0,Y=K,Z=M;  TVECTOR U,V;
      U_|(X*Y^Z);           % inner product
               K
      X*(( - 1) *Y^U_|Z + U_|Y^Z)
 
      PFORM Z=K; TVECTOR U;
      U |_ Z;               % Lie derivative
      U_|d Z + d(U_|Z)
\end{verbatim}

A \textbf{metric structure} is defined in EXCALC by specifying a set of
basis one- forms (the coframe) together with the metric. The clause WITH
METRIC can be omitted if the metric is Euclidean and the shorthand WITH
SIGNATURE $<$diagonal elements$>$ can be used in the case of a
pseudo-Euclidean metric. The splitting of a metric structure in its metric
tensor coefficients and basis one-forms is completely arbitrary including
the extremes of an orthonormal frame and a coordinate frame. 
Examples (\cite{2}) :
\begin{verbatim}
   COFRAME O(T)=D T, O X=D X
   WITH SIGNATURE -1,1;      %A Lorentz coframe;
 
   COFRAME E R=D R, E PH=D PH             %Polar coordinate
   WITH METRIC G=E R*E R+R**2*E PH*E PH;  %basis;
\end{verbatim}

The frame, dual to the coframe defined by the COFRAME command can be
introduced by FRAME $<$identifier$>$. This command causes the dual property
to be recognized, and the tangent vectors of the coordinate functions are
replaced by the frame basis vectors.

The command RIEMANNCONX is provided for calculating the connection 1 forms.
Example : calculate the connection 1-form and curvature 2-form on S(2) 
(displaying only the nonzero results) :
\begin{verbatim}
      COFRAME E TH=R*D TH,E PH=R*SIN(TH)*D PH;
      RIEMANNCONX OM;
      OM(K,-L);         %Display the connection forms;
 
        PH         PH
      NS      := (E  *COS(TH))/(SIN(TH)*R)
           TH
 
        TH            PH
      NS      := ( - E  *COS(TH))/(SIN(TH)*R)
           PH
 
      PFORM CURV(K,L)=2;
      CURV(K,-L):=D OM(K,-L) + OM(K,-M)^OM(M-L);
                                   %The curvature forms
 
          PH            TH  PH   2
      CURV      := ( - E  ^E  )/R   %it was a sphere with
              TH                     %radius R.
 
           TH         TH  PH   2
      CURV      := (E  ^E  )/R
              PH
\end{verbatim}

\section{General relativity on the computer.\\Schwarzschild solution}

The students in our Faculty of Physics are, generally speaking, well trained
in practical computer manipulations. There is no semester without at least
one course with labs in the computer room. But when we invited our students
to come in the computer room, to learn something about general relativity
with the computer, it was a general surprise, because they considered (until
now) the computer as a tool for hard \textbf{numerical} computations. They
do not know almost anything about \textbf{computer algebra} - \cite{3}.

It is not necessary to use sophisticated procedures, with large and
complicated metric statements (which are almost impossible to calculate by
hand in a civilized time of teaching) in order to convince a dedicated
student in general relativity about the utility of a computer algebra
system. It is enough to run a simple program like (\cite{4}) :
\begin{verbatim}
pform psi=0; fdomain psi=psi(r);
coframe
   o(t)     = psi           * d t,   % Schwarzschild
   o(r)     = (1/psi)       * d r,   %  metric
   o(theta) = r             * d theta,
   o(phi)   = r * sin(theta)* d phi
with signature 1,-1,-1,-1; frame e;
\end{verbatim}

to introduce a Schwarzschild type metric in spherical coordinates $%
(r,\theta,\varphi)$. This means that in classical notation the interval is 
\begin{equation}  \label{sch}
ds^2 = \Psi^2 dt^2 - \frac{1}{\Psi^2}dr^2 - r^2\left (d\theta^2 +
sin^2\theta~d\varphi^2\right )
\end{equation}
$\Psi$ being a function $\Psi=\Psi(r)$ and differ from a Minkowski one by
the a new "$unknown$" function ($\Psi = \sqrt{1+unknown(r)}$ which must be
determined from Einstein equations :
\begin{verbatim}
pform unknown=0; fdomain unknown=unknown(r);
psi := sqrt(1 + unknown);
\end{verbatim}

Now comes the most important part of the procedure : the calculation of the
components of Einstein tensor ($einstein3$) via the Riemann or Levi-Civita
connection 1-form $\Gamma^{ij}$ - $christ1$ and the curvature 2-form $R^{ij}$
- $curv2$ 
\begin{verbatim}
pform chris1(a,b)=1, curv2(a,b)=2, einstein3(a)=3;
antisymmetric chris1, curv2;
riemannconx christ1; chris1(a,b) := christ1(b,a);
curv2(a,b)      := d chris1(a,b) + chris1(-c,b) ^ chris1(a,c);
einstein3(-a)   := (1/2) * curv2(b,c) ^ #(o(-a)^o(-b)^o(-c));
\end{verbatim}

The last of the above program lines just defines the Einstein 3-form which
appears in the Einstein equations. Those who prefer the coordinate
components form (thus the Einstein tensor $G_{ij}= R_{ij}-\frac{1}{2}g_{ij}R$%
) can use the next line to ``pick-up'' these components as :
\begin{verbatim}
pform Ein(i,j)=0;
Ein(-i,-j):=e(-i)_|#einstein3(-j);
\end{verbatim}

A typical component ($G^{\phi}$) of the output of $einstein3$ reads :
\begin{verbatim}
     T  R  THETA
 - (O ^O ^O     *(@   UNKNOWN*R + 2*@ UNKNOWN))/(2*R)
                   R R               R
\end{verbatim}

or : 
\begin{eqnarray}
G^{\phi} = \left (- \frac{1}{2}\frac{\partial^2 unknown}{\partial r^2} - 
\frac{1}{r}\frac{\partial unknown}{\partial r} \right ) o^{t} \wedge o^{r}
\wedge o^{\theta}  \nonumber
\end{eqnarray}
Requiring the coefficients to vanish yields a second order differential
equation for the function $unknown$. Trying $unknown = \alpha * m/r**n$,
after using SOLVE (\cite{1}) package (i.e. the EXCALC command {\bf 
solve(einstein3(-phi),unknown);} ), we obtain $n=-1$ and :
\begin{verbatim}
unknown := - alpha * m/r;
\end{verbatim}

\noindent where "$m$" is the mass and $alpha$ a constant coefficient to be 
determined
by physical considerations (link to the Newtonian theory, for example).

Finally, evaluating the $\psi$ function ($psi := psi$), we obtain : 
\begin{eqnarray}
\psi = \sqrt{1-\frac{\alpha m}{r}}  \nonumber
\end{eqnarray}
or, in (\ref{sch}) we have 
\begin{equation}  \label{schw}
ds^2 = \left (1-\frac{\alpha m}{r}\right )dt^2 - \frac{1}{1-\frac{\alpha m}{r%
}}dr^2 - r^2\left (d\theta^2 + sin^2\theta~ d\varphi^2\right )
\end{equation}
which is the typical form of the Schwarzschild metric (\cite{4}-\cite{5})
identifying $\alpha = 
2$ by physical considerations.

From now one, it is possible to study, in a similar way more complex
situations, like Reissner-Nordstr\" om  metric (starting, of course with
the above Schwarzschild one - this example is presented in detail in 
\cite{6}). The teacher can select more exact solutions of
Einstein equations in order to complete the education of his students. Also,
algebraic programming can be used to present (in a very fast manner) the
canonical version of general relativity (\cite{7}) or the post-Newtonian
approximation (\cite{4}).

The next section is dedicated to more examples which can be 
used in the teaching process of general relativity.

\section{Other examples}

According to the literature (see, for example \cite{4} or \cite{8}) 
two of the 
homogeneous solutions of the Einstein equations with cosmological constant
are the de Sitter and anti-de Sitter metrics, having the line element 
written as :
\begin{equation}\label{sittg}
ds^2 = e^{n g}dt^2 - e^{k f}dx^2 - e^{n g +k f}(dy^2 +dz^2)
\end{equation}
where $f$ and $g$ are two unknown functions of $x$ and $t$ variables which
we shall determine by imposing that the vacuum Einstein equations with 
cosmological constant to be fulfilled :
\begin{equation}\label{einscos}
G_{ij}=R_{ij}-\frac{1}{2}g_{ij}R - \Lambda g_{ij} =0
\end{equation}
Also in the above equation (\ref{sittg}) $n$ and $k$ are two constants 
introduced in 
order to control the two solutions : for $n=0$ and $k=1$ we shall obtain
the de Sitter solution and for $n=1$ and $k=1$ we have the anti-de Sitter
solution. Thus we can use a program sequence for introducing this metric as :
\begin{verbatim}
pform f=0,g=0; fdomain f=f(x,t),g=g(x,t);
coframe
   o(t)     =  exp(n*g/2)          * d t,   
   o(r)     =  exp(k*f/2)          * d x,   
   o(theta) =  exp((n*g+k*f)/2)    * d y,
   o(phi)   =  exp((n*g+k*f)/2)    * d z
with signature 1,-1,-1,-1; frame e;
\end{verbatim}
Now we must run the sequences which calculates the 
Einstein tensor $G_{ij}$ components with cosmological, namely :
\begin{verbatim}
riemannconx chris;
pform chris1(a,b)=1;
pform curv2(a,b)=2;
pform einstein3(a)=3;
antisymmetric chris1;
antisymmetric  curv2;
chris1(a,b):=chris(b,a);
curv2(a,b):= d chris1(a,b)+chris1(-c,b)^chris1(a,c);
einstein3(-a):=(1/2)*curv2(b,c)^ #(o(-b)^o(-c)^o(-a))-Lam*#o(-a);
pform Ein(i,j)=0;
Ein(-i,-j):=e(-i)_|(#einstein3(-j));
\end{verbatim}
where ${\bf Lam}$ represents the cosmologic constant $\Lambda$ and 
{\bf Ein(-i,-j)} represents the components of the Einstein tensor $G_{ij}$ 
with cosmological constant in  coordinate frame. A typical output is, for
example :
\begin{verbatim}
                 F*K + G*N          F*K    2  2
EIN    := ( - 4*E         *LAM + 3*E   *@ F *K
   T T                                   T

                 F*K                F*K    2  2
            + 4*E   *@ F*@ G*K*N + E   *@ G *N
                      T   T              T

                 G*N            G*N    2  2
            - 4*E   *@   F*K - E   *@ F *K
                      X X            X

                 G*N                  G*N
            - 4*E   *@ F*@ G*K*N - 4*E   *@   G*N
                      X   X                X X

                 G*N    2  2      F*K + G*N
            - 3*E   *@ G *N )/(4*E         )
                      X
\end{verbatim}
which means, in ``normal'' form :
\begin{eqnarray}
G_{tt} = e^{-ng}\left [ \frac{3}{4}k^2 \left ( \frac{\partial f}{\partial t}
\right )^2 +
kn\frac{\partial f}{\partial t}\frac{\partial g}{\partial t} +
\frac{1}{4}n^2\left (\frac{\partial g}{\partial t}\right )^2\right ] -
~~~~~~~~~~~~~~~~~~~~~~~~~~~~~~\nonumber\\
~~~~~~~~~~~~~~~e^{-kf}\left [k\frac{\partial^2 f}{\partial x^2} +\frac{1}{4} k^2 \left (
\frac{\partial f}{\partial x}\right )^2 + kn \frac{\partial f}{\partial x}
\frac{
\partial g}{\partial x} +n \frac{\partial^2 g}{\partial x^2} +
\frac{3}{4}n^2 \left (\frac{\partial g}{\partial x}\right )^2\right ] 
-\Lambda\nonumber
\end{eqnarray}
Now assigning to the parameters the values $n=0$ and $k=1$ the above component
of the Einstein tensor) and the corresponding Einstein equation) will be :
\begin{eqnarray}
G_{tt} = \frac{3}{4}\left (\frac{\partial f}{\partial t}\right )^2-
e^{-f}\left [\frac{\partial^2 f}{\partial x^2}+\frac{1}{4}\left (\frac{\partial f}
{\partial x}\right )^2\right ] - \Lambda = 0\nonumber
\end{eqnarray}
By inspecting this equation (\cite{9}) we can observe that his general 
solution  can be written as $f(x,t)= f1(t) + f2(x)$ where
\begin{eqnarray}
\left (\frac{d f1(t)}{d t}\right )^2 = \frac{1}{3}
C_1 e^{-f1(t)}+\frac{4}{3}\Lambda\hbox{~~~and~~~}\nonumber\\
\frac{d^2 f2(x)}{d x^2}=\frac{1}{4}C_1 e^{f2(x)}-\frac{1}{4}\left (
\frac{d f2(x)}{d x}\right )^2\nonumber
\end{eqnarray}
where $C_1$ is a constant.
Thus we can choose $f(x,t)=f1(t)$ such that the above equation becomes :
\begin{verbatim}
pform f1=0;fdomain f1=f1(t);f:=f1;
Ein(-t,-t);
       2
 3*@ F1  - 4*LAM
    T
-----------------
        4
\end{verbatim}
as computer output and
\begin{eqnarray}
G_{tt}=\frac{3}{4}\left (\frac{d f1}{d t}\right )^2 - \Lambda=0\nonumber
\end{eqnarray}
This equation has an obvious solution of the form :
\begin{eqnarray}
f1=f= \frac{2}{3}\sqrt{3 \Lambda}t +C_1\nonumber
\end{eqnarray}
which fulfill the Einstein equations with cosmological constant (all the
components os {\bf Ein(-i,-j)} 0-form are zero). Then the line element
becomes :
\begin{eqnarray}
ds^2 =  dt^2 -e^{\frac{2}{3}\sqrt{3\Lambda}t+C_1}\left (dx^2 +dy^2 +
dz^2\right )
\nonumber
\end{eqnarray}
which is, of course the well-known De Sitter solution. For the anti-De Sitter
solution we must choose the parameters as $n=1$ and $k=0$. Thus the
interesting Einstein tensor components become :
\begin{verbatim}
n:=1;k:=0;
Ein(-t,-t);

       G            G    2      G          2
  - 4*E *@   G - 3*E *@ G  - 4*E *LAM + @ G
          X X          X                 T
---------------------------------------------
                       G
                    4*E

Ein(-x,-x);

    G    2      G                    2
 3*E *@ G  + 4*E *LAM - 4*@   G - @ G
       X                   T T     T
---------------------------------------
                    G
                 4*E
\end{verbatim}
as computer output, or in ``normal'' transcript :
\begin{eqnarray}
G_{tt} = \frac{1}{4}e^{-g}\left (\frac{\partial g}{\partial t}\right )^2-
\frac{\partial^2 g}{\partial x^2} -\frac{3}{4}\left ( 
\frac{\partial g}{\partial x}\right )^2 -\Lambda
=0\nonumber\\
G_{xx} = \frac{3}{4} \left ( \frac{\partial g}{\partial x}\right )^2 -
e^{-g}\left [\frac{\partial^2 g}{\partial t^2} +\frac{1}{4}\left (
\frac{\partial g}{\partial t}\right )^2\right ] +\Lambda =0\nonumber
\end{eqnarray}
Solving, in a similar way as in the De Sitter case, the second of the above
Einstein equations, we have to impose $g(x,t):=g1(x)$. Thus we obtain a
solution as :
\begin{eqnarray}
g=g1= \frac{2}{3}\sqrt{-3\Lambda}x +C_2\nonumber
\end{eqnarray}
where $C_2$ is a new constant. We have the line element as
\begin{eqnarray}
ds^2 = -dx^2 + e^{\frac{2}{3}\sqrt{-3\Lambda}x+C_1}\left (
dt^2 - dy^2 - dz^2\right )\nonumber
\end{eqnarray}
which is the anti-De Sitter solution. 

The REDUCE+EXCALC procedures we presented here are also suitable for a
series of other applications during the relativity lab classroom (...). For
example the teacher can propose to the students to check if other metrics
fulfill the Einstein equations. They can use the above program sequences with
minor modifications. As an exercise we propose here the Ozsvath metric 
(\cite{8})
\begin{eqnarray}
ds^2 = dx^2 + e^{2\sqrt{-\Lambda/3}x}\left ( dy^2 + 2 du dv\right ) +
f e^{-\sqrt{-\Lambda/3}x}\left ( -2 \sqrt{2} dy + f e^{-3\sqrt{-
\Lambda/3}x}dv\right )dv\nonumber
\end{eqnarray}
which is also an homogeneous solution of vacuum Einstein equations with
cosmological term. Here a special care must be taken to the changing
of the signature of the metric and to the nondiagonal terms of the metric.

\end{document}